\documentclass[final]{svjour2}
\usepackage{graphicx}
\usepackage{rotating}
\usepackage{amssymb}
\usepackage{amsmath}
\usepackage{bm}
\usepackage[square,numbers]{natbib}
\makeatletter
\journalname{Journal of Low Temperature Physics}

\def\tdp{\tau}

\def\ol{\omega_{\rm L}}
\def\oe{\omega_{\rm m}}
\def\op{\omega^\prime}
\def\tp{{\tdp}^\prime}

\def\sm{{S_\perp^{(-)}}^2}
\def\sp{{S_\parallel^{(+)}}^2}

\def\kpa{{k_\parallel}}

\def\vF{v_{\rm F}}

\begin{document}

\renewcommand{\topfraction}{0.95}
\renewcommand{\textfraction}{0.05}

\newcommand{\hdblarrow}{H\makebox[0.9ex][l]{$\downdownarrows$}-}
\title{Microkelvin thermometry with Bose-Einstein condensates of magnons and applications to studies of the AB interface in superfluid $^3$He}

\author{P.J.~Heikkinen \and S.~Autti \and V.B.~Eltsov \and R.P.~Haley \and V.V.~Zavjalov}

\institute{
P.J. Heikkinen \and S. Autti \and V.B. Eltsov \and V.V. Zavjalov \at
O.V. Lounasmaa Laboratory, Aalto University, P.O. Box 15100, 00076 AALTO, Finland\\
\email{petri.heikkinen@aalto.fi}\\
\\
R.P. Haley \at
Department of Physics, Lancaster University, Lancaster, LA1 4YB, UK\\
}
\date{Received: date / Accepted: date}

\maketitle

\begin{abstract}

  Coherent precession of trapped Bose-Einstein condensates of magnons
  is a sensitive probe for magnetic relaxation processes in
  superfluid $^3$He-B down to the lowest achievable temperatures. We use the dependence of the relaxation rate on the density of thermal
  quasiparticles to implement thermometry in $^3$He-B at temperatures below
  $300\,\mu$K. Unlike popular vibrating wire or quartz tuning fork based
  thermometers, magnon condensates allow for contactless temperature
  measurement and make possible an independent \textit{in situ}
  determination of the residual zero-temperature relaxation provided by the
  radiation damping. We use this magnon-condensate-based thermometry to
  study the thermal impedance of the interface between A and B phases of
  superfluid $^3$He. The magnon condensate is also a sensitive probe of the
  orbital order-parameter texture. This has allowed us to observe for the first
  time the non-thermal signature of the annihilation of two AB interfaces.


\end{abstract}

\section{Introduction}

The superfluid phases of $^3$He are unique examples of topological Fermi superfluids~\cite{volovik-droplet} which can be experimentally studied at temperatures much below the transition temperature $T_{\mathrm{c}}$: there are technical difficulties for cooling superfluid atomic Fermi gases into this temperature regime~\cite{Fermi-gas-theory_RMP80} and the neutron superfluid in neutron stars~\cite{neutron-star_science} is not available for laboratory investigation. Measurements of ultra-low temperatures in superfluid $^3$He present an experimental challenge, though, since the thermal link between the liquid and a solid wall rapidly deteriorates with decreasing temperatures~\cite{parpia_PRB32}, making the use of external thermometers more and more difficult. Reliable determination of temperatures below about $0.25T_{\mathrm{c}}$ requires probing of the properties of the liquid itself.

This goal can be achieved with a technique which was pioneered in Lancaster and is based on measurements of the damping of an oscillating object immersed in bulk superfluid $^3$He~\cite{vibrating-wire_JLTP62}. Originally, loops made from thin superconducting wire were used as oscillators. Later, quartz tuning forks also became popular~\cite{fork_JLTP}. The damping of an oscillating object turns out to be proportional to the density of thermal quasiparticles. In the B phase, which we discuss in this paper, the density rapidly decreases at low temperatures as $\propto \exp(-\Delta/k_{\mathrm{B}} T)$, where $\Delta$ is the superfluid energy gap.

The proportionality coefficient between this exponential factor and the measured value of damping can be estimated from the geometry of the oscillating object~\cite{damping-lanc_JLTP157}, and then further calibrated by comparing to another thermometer at temperatures of
$\lesssim 0.3T_{\mathrm{c}}$. As a reference thermometer one can use, for example, platinum NMR~\cite{VWR_grenoble}, the $^3$He melting
curve~\cite{fork_JLTP, Igor_JLTP126}, or NMR on superfluid $^3$He (see below). All oscillating objects used so far as thermometers have intrinsic damping in addition to the damping caused by the $^3$He. This extra contribution limits thermometer sensitivity at the lowest temperatures and its reliable experimental determination is difficult.

Another traditional way to probe superfluid $^3$He directly is nuclear magnetic resonance. In fact, thermometry based on the NMR properties of $^3$He has been used for decades and predates the invention of vibrating-wire thermometers. The width of the NMR spectrum of superfluid $^3$He or the frequency shifts of characteristic spectral features away from the Larmor frequency are expressed through the Leggett frequency $\Omega_{\mathrm{B}}(P,T)$, which characterizes the strength of the spin-orbit interaction. Rapid variation of $\Omega_{\mathrm{B}}(T)$ below $T_{\mathrm{c}}$ allows for convenient temperature measurements. However, the temperature dependence of $\Omega_{\mathrm{B}}(T)$ saturates at $T \sim 0.25T_{\mathrm{c}}$ and so this method loses sensitivity at lower temperatures~\cite{ahonen_JLTP25}.

In a rotating sample of $^3$He-B in a metastable vortex-free state the NMR spectrum includes a characteristic peak, the so-called counterflow peak~\cite{korhonen_PRL65}, with an amplitude which depends, in particular, on the anisotropy of the superfluid density in magnetic field. This anisotropy rapidly vanishes at low temperatures and the decreasing height of the counterflow peak then becomes a sensitive temperature probe. It becomes especially sensitive in a narrow temperature interval where a transition in the orbital texture causes the complete disappearance of the counterflow peak~\cite{texture-rob_JLTP163} --- in a sense similar to superconducting transition-edge temperature sensors. However, this method is applicable only in very specialized situations.

In this paper we describe a more general approach to temperature measurement in ultra-cold superfluid $^3$He-B from its NMR response. This method is based on the measurement of the relaxation rate of trapped Bose-Einstein condensates (BEC) of magnon quasiparticles~\cite{bunvol_Qball,bunvol_review}, a state originally referred as the ``persistent induction signal (PIS)''~\cite{PIS_PRL69} and later as the ``persistent precessing domain (PPD)''~\cite{PPD_JLTP134}. The magnon condensate in a trap is manifested by a spontaneous long-lived coherent precession of the magnetization in an external magnetic field.

A three-dimensional magnon trap can be formed by the combined effect of the orbital order-parameter texture, which is controlled by the spin-orbit interaction energy, and a spatially varying magnetic field controlling the Zeeman energy. In a cylindrical sample of $^3$He-B with an axially oriented magnetic field the traditional approach is to use the flare-out orbital texture for radial trapping of the condensate close to the sample axis, and to create a minimum in the applied magnetic field for axial trapping close to the center of the NMR pick-up coil, see Fig.~\ref{fig:meas_setup}.

When magnons with energies larger than the ground level in the trap are pumped into the system, for example by an rf pulse of suitable frequency and amplitude, they quickly relax to the ground state where a spontaneously-coherent magnon BEC is formed~\cite{self-trapping_PRL}. The coherently precessing magnetization of the condensate induces a voltage in the pick-up coil. The decay rate of this signal measures the loss of magnons from the trap. In a trap which is well isolated from the sample walls two loss mechanisms have been identified: spin diffusion through the normal component~\cite{relax_lancaster, magnon_relax_JLTP} and
radiation damping owing to the interaction with the pick-up circuit~\cite{magnon_relax_JLTP}.

The spin transport via the normal component in the ballistic regime~\cite{Einzel_JLTP84}, which is still customarily called the spin
diffusion, results in a fast variation of the relaxation rate with temperature. The central property of the spin transport in this
regime~\cite{Leggett-PRL,Leggett-long,Bunkov_etal_PRL65,mark-mukh} is the Leggett-Rice effect associated with the existence of the Landau molecular field. As a result, the effective spin diffusion coefficient $D$ acquires a fast temperature dependence similar to the damping of the oscillating object, $D \propto \exp(-\Delta/k_{\mathrm{B}} T)$ (see appendix). This is the basis for the temperature measurements. The geometry of the condensate, which figures essentially into the relaxation rate, is determined usually by a nearly harmonic trap. The shape of the trapping potential (and thus the spatial distribution of the magnetization) can also be resolved, if needed, using spectroscopy of the excited levels in the trap~\cite{magnon_relax_JLTP}.

Radiation damping~\cite{raddamp} provides an additional approximately temperature-independent contribution to the magnetic relaxation, which thus persists towards zero temperature, similar to the intrinsic damping in mechanical oscillators. However, unlike in mechanical systems, in the magnon condensate this additional damping can be determined in situ by measuring the frequency dependence of the relaxation. This allows for fully self-calibrating thermometry if one assumes that the theoretical value of the spin-diffusion
coefficient is reliable. If both magnon-condensate and mechanical-oscillator-based thermometers are thermally linked within the
same volume of $^3$He-B, the condensate can be used to independently measure the intrinsic damping in the mechanical oscillator, as shown below. In this case the temperature calibration of the mechanical thermometer, if known, can also be used to calibrate the condensate thermometer without knowledge of the spin-diffusion coefficient.

We have applied this magnon-condensate-based thermometry to study the thermal properties of the interface between $^3$He-A and $^3$He-B. Superfluid coherence is preserved across the AB interface, but the order parameter changes rapidly on the length scale of the superfluid coherence length. This presents a barrier for the motion of quasiparticles and thus leads to a thermal impedance across the interface. This impedance was measured in Lancaster using vibrating-wire thermometers and heaters~\cite{AB-brane_JLTP}. We have performed similar measurements at lower temperatures using the magnon BEC as one of the thermometers. The thermal impedance in our case is found to be larger than in the earlier measurements, as expected for quasiparticles with lower thermal energies.

In Lancaster it was also found that the thermal resistance of the B-phase column increased after the A phase had been created and subsequently removed in some section of a long cylinder by sweeping the magnetic field first up and then down, so that two AB interfaces were annihilated~\cite{AB-brane}. We have not confirmed this effect. Instead, we have observed a different signature from the interface annihilation. The precession frequency of the magnon condensate is a probe for the orbital order-parameter texture in $^3$He-B. Immediately after the annihilation event we observe fluctuations in the precession frequency. This demonstrates for the first time that some non-trivial orbital dynamics are triggered in $^3$He-B in the presence of a rapidly propagating AB interface.

\section{Experimental setup}

\begin{figure}[t]
\centering
\flushleft
\includegraphics[width=0.99\textwidth]{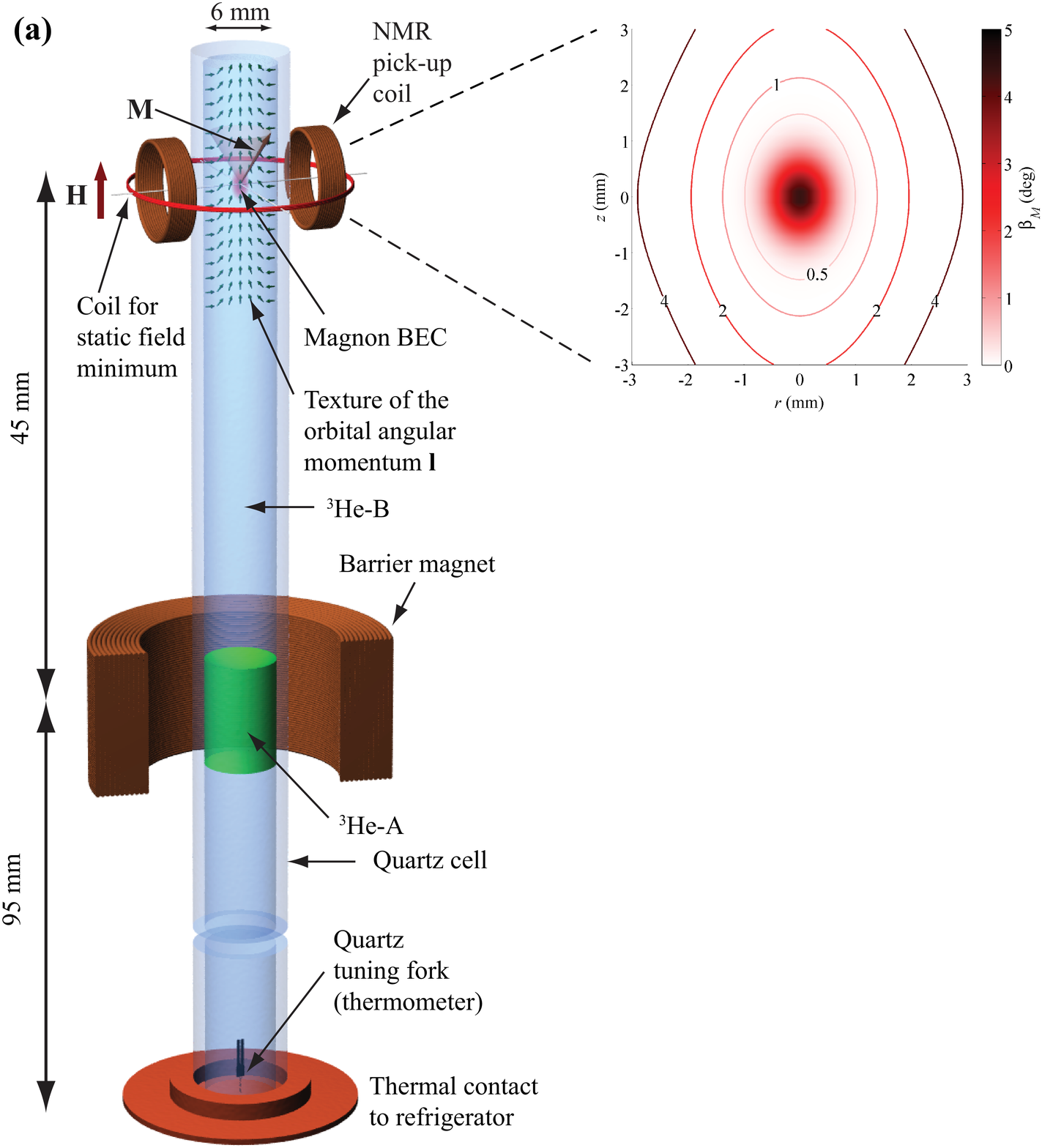}
\hspace{-1.3cm}
\llap{{
\includegraphics[width=0.32\textwidth]{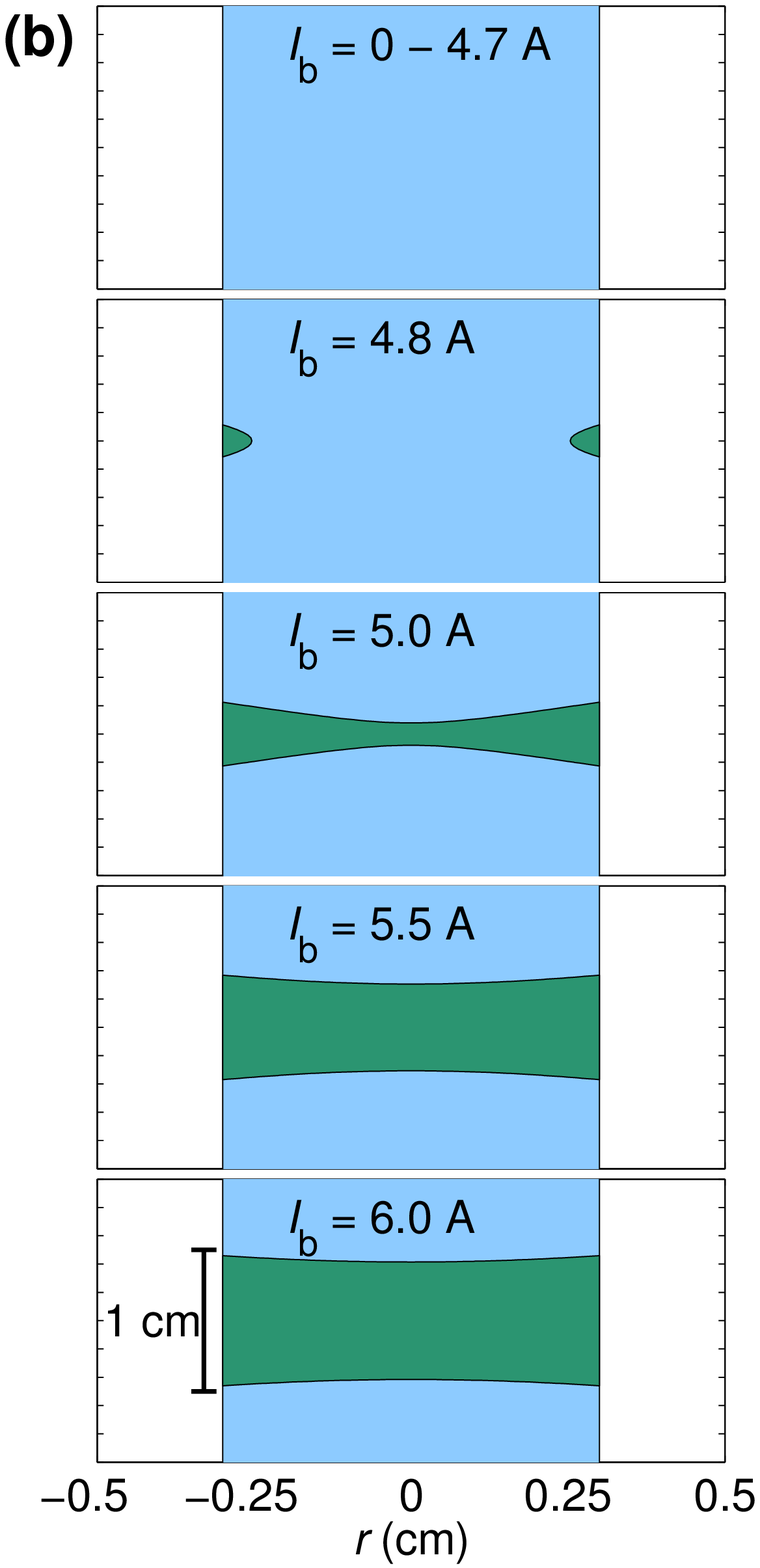}
}}
\caption{\label{fig:meas_setup}(Color online) The setup used for the experiments with the magnon BEC and the AB interface. \textbf{(a)} The vertical sample cylinder is filled with superfluid $^3$He. In the top part an NMR spectrometer is installed. It is equipped with an additional coil to provide a minimum in the polarizing magnetic field $H$ for trapping the magnon BEC in the axial direction. Radial trapping is provided by the orbital order-parameter texture (shown with small arrows). The contour map of the total potential (in kHz) and the respective distribution of the tipping angle $\beta_M$ of the condensate magnetization in the ground state of the trap are shown in the plot on the top right. The barrier magnet can be used to create a slab of $^3$He-A in the middle of the sample, while the rest remains in the B phase. Close to the bottom a quartz tuning fork is installed, which is used as a thermometer. At the bottom the sample is connected to the sintered heat exchanger volume of the nuclear cooling stage. \textbf{(b)}~Calculated profiles of the A-phase layer for different currents $I_{\mathrm{b}}$ in the barrier magnet at $T=0.13T_{\mathrm{c}}$, $P=4.1\,$bar. Data for the equilibrium B$\rightarrow$A transition field is taken from Ref.~\cite{hahn_thesis}. The surface tension and the wetting angle of the AB interface are not taken into account.}
\end{figure}

The superfluid $^3$He sample fills a 15\,cm long cylindrical container which is made from fused quartz, Fig.~\ref{fig:meas_setup}. The pick-up coil of the NMR spectrometer is placed 1\,cm below the upper end of the tube. The coil is part of a tuned tank circuit with a Q factor of about 130. The capacitance in the tank circuit is placed at mixing chamber temperature and the tank circuit is connected to a cold preamplifier at the 4K flange. To change the resonance frequency of the tank circuit we use a bank of capacitors where the capacitance can be selected with latching relays. In all measurements in this paper which involve the A phase the strength of the static magnetic field $H$ is 26\,mT with an inhomogeneity $\Delta H / H \approx 4\cdot 10^{-4}$. This corresponds to a Larmor frequency of $f_{\mathrm{L}}=\omega_{\mathrm{L}}/2\pi=833$\,kHz. The measurements are performed at 4.1\,bar pressure.

At this pressure B phase is stable at low magnetic fields. With the help of the barrier magnet the field can be increased above the critical field $H_{\mathrm{AB}}$ for the B$\rightarrow$A transition and a layer of $^3$He-A can be stabilized at a distance of about 4\,cm from the NMR spectrometer. Under the conditions of our experiment $H_{\mathrm{AB}} = 0.465\,$T~\cite{hahn_thesis}. This field is reached first at the cylindrical boundary of the sample when the current in the barrier magnet is increased to $I_{\mathrm{b}}=4.8\,$A. If $I_{\mathrm{b}}$ is further increased, the equilibrium volume occupied by the A phase increases as shown in Fig.~\ref{fig:meas_setup}b.

The trapped magnon condensates are created as described in detail in Refs.~\cite{self-trapping_PRL,magnon_relax_JLTP}. The magnetic part of the trapping potential is provided by an additional pinch coil which creates a field in the opposite direction to that of the main polarizing field. Thus the total field and the Zeeman energy $F_{\mathrm{Z}}=\hbar \omega_{\mathrm{L}}|\Psi|^2=\hbar\gamma H|\Psi|^2$, where $\gamma$ is the absolute value of the gyromagnetic ratio and $\Psi$ is the wave function of the magnon condensate, have a shallow local minimum in the axial direction inside the NMR measurement volume. Radial trapping is provided via the spin-orbit interaction energy by the flare-out texture of the orbital anisotropy axis of the order parameter:
\begin{equation}
 \label{eq:spin-orbit}
 F_{\mathrm{so}} = \frac45 \hbar \frac{\Omega_{\mathrm{B}}^2}{\omega_{\rm L}}\sin^2 \frac{\beta_l(r,z)}{2}\,|\Psi|^2,
\end{equation}
where the strength of the interaction is characterized by the Leggett frequency $\Omega_{\mathrm{B}}$ and the spatial dependence comes from the variation of the angle $\beta_l$ between the orbital anisotropy axis $\hat{\mathbf{l}}$ and the magnetic field. The role of these two energies separately in enabling the coherent spin precession is studied in Ref.~\cite{kupka_skyba} for a field minimum and Ref.~\cite{bunvol_Qball} for a textural confinement. Together $F_{\mathrm{Z}}$ and $F_{\mathrm{so}}$ form a three-dimensional trapping potential in which the magnon condensate is well isolated from the sample boundaries.

A remarkable property of the textural trapping potential is that it can be modified by the precessing magnetization~\cite{bunvol_Qball,self-trapping_PRL}. In this paper we, however, discuss condensates with a sufficiently small number of magnons, so that this self-modification of the trapping potential can be neglected. In this case the trapping potential is close to harmonic at the bottom of the trap. Thus the lowest energy levels obey the familiar harmonic spectrum: $\omega_{n_r n_z}=\omega_{\mathrm{L}}(r=0, z=0) + \omega_r(n_r+1)+\omega_z(n_z+1/2)$, where $\omega_r$ and $\omega_z$ are the radial and axial trapping frequencies, and the quantum numbers $n_r$ and $n_z$ can take only even values. A simple way to experimentally determine $\omega_r$ and $\omega_z$ is to measure the difference in the precession frequencies between a few lowest-energy levels in the trap using continuous-wave NMR, as described in Ref.~\cite{magnon_relax_JLTP}.

After a rapid transition from the normal to the superfluid state one often finds that the texture has some defects, such as $\hat{\mathbf{n}}$ solitons~\cite{n-soliton}. In experiments we have found that these defects change the trapping potential and the relaxation properties in an ill-defined manner, making measurements and calculations impracticable. To recover a defect-free texture and a stable potential we have used the fact that spin precession at large tipping angles forces the texture to become uniform. Thus sweeping the NMR absorption line with a sufficiently large excitation amplitude usually ``cleans'' the texture, resulting in a well-defined trapping potential.

The temperature at the lower end of the sample container is measured with a quartz tuning fork~\cite{fork_JLTP}. The fork is calibrated in the temperature range $(0.3-0.35)T_{\mathrm{c}}$ using frequency-shift-based NMR thermometry with a second NMR spectrometer (not shown in Fig.~\ref{fig:meas_setup}), located in the lower part of the sample below the A-phase region. (See, for example, Fig.~1 in Ref.~\cite{magnon_relax_JLTP}.) The fork resonance width towards low temperatures is extrapolated using the weak-coupling-plus energy gap~\cite{wcp-gap}. The heat leak into the sample cylinder was measured to be 12\,pW in earlier experiments where the sample cylinder was separated from the sintered heat exchanger volume by a plate with a small orifice~\cite{hosio_thermal_PRB84}, using ``black-body radiator'' techniques~\cite{blackbody_lanc}.

\section{Magnon BEC thermometry}

\begin{figure}
\centerline{\includegraphics[width=\textwidth]{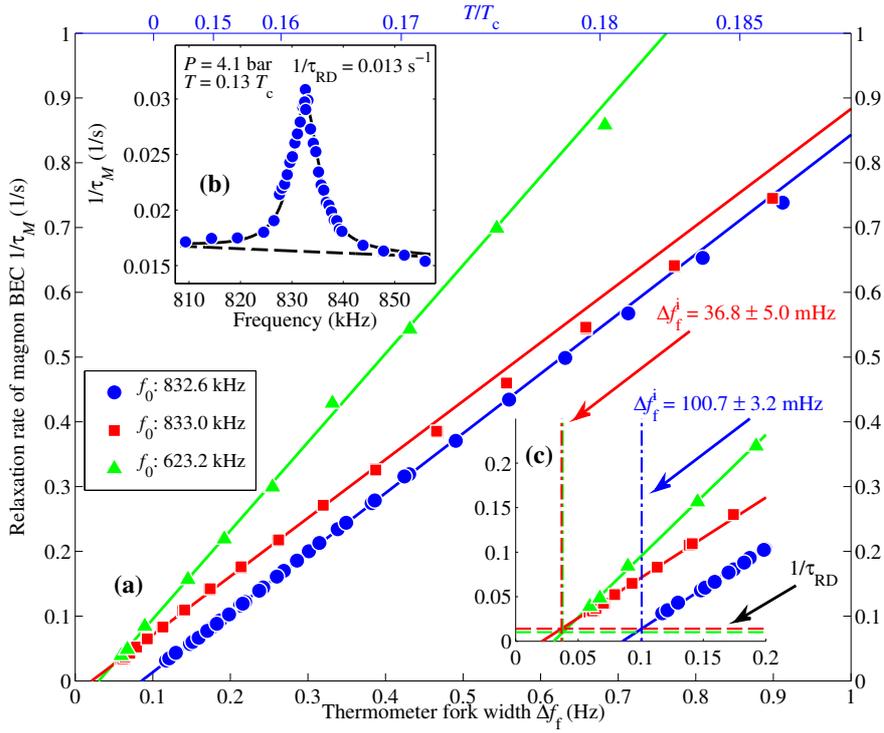}}
\caption{\label{fig:t_calibration} (Color online) Relaxation rate $\tau_M^{-1}$ of the magnon BEC versus reduced temperature $T/T_{\mathrm{c}}$ (top axis) and density of thermal quasiparticles $[\propto \exp (-\Delta / k_{\mathrm{B}}T)]$ as measured from the width $\Delta f_{\mathrm{f}}$ of a quartz tuning fork resonance (bottom axis). (\textbf{a}) The dependence of $\tau_M^{-1}$ on $\Delta f_{\mathrm{f}}$ (symbols) is nearly linear (lines) since at low $T$ both the spin diffusion coefficient $D$ and $\Delta f_{\mathrm{f}}$ depend on the density of quasiparticles, i.e., exponentially on temperature. The linear fits include only points below $T = 0.175T_{\mathrm{c}}$ since the deviation of $D$ from the exponential temperature dependence becomes evident at higher temperatures. The slope of this dependence increases with decreasing precession frequency $f_0=\omega_{00}/2\pi$ according to Eq.~(\ref{eq:tau_M}). The zero intercept of the lines depends in particular on the intrinsic width $\Delta f_{\mathrm{f}}^{\mathrm{i}}$ of the fork as seen from the measurement with a deteriorated fork (see text) and an increased $\Delta f_{\mathrm{f}}^{\mathrm{i}}$ (filled circles). The temperature scale can be restored even for this data (top axis) if the radiation damping is measured. (\textbf{b}) Measurement of $\tau_M^{-1}$ versus frequency of precession (circles) reveals a contribution from the radiation damping which is fitted using the Lorentzian resonance response of the pick-up tank circuit (solid line) above the spin-diffusion background (dashed line). (\textbf{c}) Zoom of plot (a) showing the two different measured values of $\tau_{\mathrm{RD}}^{-1}$ as horizontal lines and the extracted values of the intrinsic fork width $\Delta f_{\mathrm{f}}^{\mathrm{i}}$ as vertical lines. For all data in the figure $I_{\mathrm{min}}=2.0\,$A, $\omega_r/2\pi= 200\,$Hz and $\omega_z/2\pi=53\,$Hz at $f_0 = 833\,$kHz, $\omega_r/2\pi= 297\,$Hz and $\omega_z/2\pi=61\,$Hz at $f_0 = 623\,$kHz. }
\end{figure}

To create the magnon condensate one first pumps non-equilibrium magnons into the system. This is achieved by applying a transverse rf magnetic field of frequency $\omega_{\mathrm{rf}}$ with the NMR coil, either continuously or as a pulse. If the spectrum of the excitation covers some excited levels $\omega_{n_r n_z}$ in the magnon trap, these levels will be filled with magnons. Magnons from the excited levels relax towards the ground state, where spontaneously coherent precession of the magnetization $\mathbf{M}$ at a frequency $\omega_{00}<\omega_{\mathrm{rf}}$ emerges with an approximately common phase that is unrelated to that of the rf excitation field. This state is known as a Bose-Einstein condensate of magnons; the method of creating it by pumping magnons at higher frequencies is called off-resonant excitation~\cite{off-reson_PRL, self-trapping_PRL}. Experimentally the off-resonant excitation is a convenient tool since separation of the excitation $\omega_{\mathrm{rf}}$ and detection $\omega_{00}$ frequencies allows us to use the same coil for both purposes without interference. Additionally, the decaying condensate can be quickly refilled by off-resonant excitation pulses without destroying its coherence. We usually use the first excited radial state (2,0) for off-resonant excitation.

The density of magnons is characterized by the tipping angle $\beta_M$ of the magnetization ${\mathbf{M}}$ away from its equilibrium direction along the magnetic field ${\mathbf{H}}$. The induction signal in the pick-up coil is proportional to the transverse magnetization of the condensate $M_\perp = \chi H \int \sin\beta_M dV$. As the condensate decays and $\beta_M$ decreases, the induction signal in the pick-up coil relaxes towards zero. When the density of magnons in the condensate is sufficiently small and self-modification of the orbital texture can be neglected, roughly corresponding to $\beta_M \lesssim 5^\circ$, the decay is exponential with a time constant $\tau_M$: $M_\perp = M_0 \exp(-t/\tau_M)$. It is the value of $\tau_M$ which we use for thermometry. At small $\beta_M$ both spin diffusion and radiation damping lead to exponential relaxation of the magnetization with time constants $\tau_{\mathrm{SD}}$ and $\tau_{\mathrm{RD}}$, respectively~\cite{magnon_relax_JLTP}. The total relaxation rate is thus a sum of the two contributions: $\tau_M^{-1} = \tau_{\mathrm{SD}}^{-1} + \tau_{\mathrm{RD}}^{-1}$.

\textit{Spin diffusion} causes dissipation of the energy of the condensate~\cite{Fomin_JETP30}:
\begin{equation}
 	\label{eq:E_diss}
	\frac{dE}{dt}=-\frac{D\gamma^2}{\chi}\left\langle \frac{\partial S_i}{\partial x_j} \frac{\partial S_i}{\partial x_j}\right\rangle,
\end{equation}
where $\chi$ is the magnetic susceptibility, $D$ is the transverse component of the spin-diffusion tensor, and indices $i$ and $j$ correspond to spin components in Cartesian coordinates. The brackets denote averaging over the volume. In the Leggett-Rice regime local gradients can be used in Eq.~(\ref{eq:E_diss}) provided that the characteristic length scale $l_{\mathrm{m}} = v_{\mathrm{g}}/\omega_{\mathrm{m}}$ is much smaller than the scale of the magnetization inhomogeneity~\cite{Leggett-long}. Here $v_{\mathrm{g}}$ is the group velocity of thermal quasiparticles and $\omega_{\mathrm{m}}/\gamma$ is the magnitude of the Landau molecular field. In our conditions $l_{\mathrm{m}} \lesssim 0.1\,$mm, while the size of the condensate $\approx 1\,$mm sets the scale of variation of the magnetization and thus this condition is satisfied. When the minimum of the magnetic field is sufficiently shallow and the relaxation is small, the inhomogeneity of the phase of spin precession in the condensate can be ignored and we get
\begin{equation}
 	\label{eq:spin_avr}
	\left\langle \frac{\partial S_i}{\partial x_j} \frac{\partial S_i}{\partial x_j}\right\rangle = \frac{(\omega_{\mathrm{L}}\chi)^2}{\gamma^4}\int(\nabla\beta_M)^2dV,
\end{equation}
where the integral is over volume $V$.

In the experiments the decay of magnetization is measured instead of energy dissipation. To connect the transverse magnetization $M_\perp$ and the energy of condensate $E$, we assume that relaxation is sufficiently slow so that the superfluid spin currents within the condensate always support it in a quasi-equilibrium state where $M_\perp$ and $E$ can be considered stationary. In this case there exists a well-defined relation $M_\perp(E)$, which allows to calculate the relaxation rate $\tau_{\mathrm{SD}}^{-1}$ from
Eqs. (\ref{eq:E_diss}) and (\ref{eq:spin_avr}):
\begin{equation}
    \label{eq:tau_SD}
    \tau_{\mathrm{SD}}^{-1}=\frac{\omega_{\mathrm{L}}^2\chi}{\gamma^2}\frac{D}{M_\perp}\frac{dM_\perp}{dE}
    \left[\int(\nabla\beta_M)^2 dV\right].
\end{equation}
This equation can be used to convert measured values of $\tau_{\mathrm{SD}}$ to temperature if the temperature dependence of $D$ and the geometry of the condensate, which determines $dM_\perp/dE$ and $\nabla\beta_M$, are known. The most thorough theoretical calculation of $D(T,P)$ which includes also the ballistic regime, which we are interested in, was performed by D.~Einzel~\cite{Einzel_JLTP84}. The final result, expressed by Eq.~(108) in Ref.~\cite{Einzel_JLTP84}, is complicated, but describes the measured values well~\cite{Bunkov_etal_PRL65,magnon_relax_JLTP}. An alternative calculation of $D$ for all temperatures is found in Ref.~\cite{mark-mukh}.

For the stationary case the Leggett equations of spin dynamics can be converted to the Schr\"{o}dinger equation for the transverse spin-wave wavefunction $\Psi$~\cite{VW-Seq}. We use this equation in the form \cite{hakonen_JLTP76}
\begin{equation}
	\label{eq:spinwave_schrode}
	\left[-\frac{24}{65}\xi^2_{\mathrm{D}}\nabla^2 +
     \frac{\omega_{\mathrm{L}}^2(r,z)}{\Omega_B^2} +
     \frac45 \sin^2 \frac{\beta_l(r,z)}{2}\right]\Psi =
     \frac{\omega_{n_r n_z}\omega_{\mathrm{L}}(0,0)}{\Omega_B^2}\Psi,
\end{equation}
where $\xi_D$ is the dipolar length. Using the ground-state ($n_r=n_z=0$) eigenfunction of this equation one then calculates $\beta_M$ from $|\Psi|^2 = \chi
H(1-\cos\beta_M)/\gamma \hbar$, $M_\perp=\chi H \int\sin\beta_M dV$, $E = \int F_{\mathrm{Z}} dV$ (since the contribution of the spin-orbit interaction to the energy of the condensate is negligible), and finally the relation between  $\tau_{\mathrm{SD}}$ and $D$ in Eq.~(\ref{eq:tau_SD}). The situation is significantly simplified in the axially symmetric harmonic trap~\cite{relax_lancaster}, where
the Eq.~(\ref{eq:tau_SD}) becomes
\begin{equation}
\tau_{\mathrm{SD}}^{-1} = \frac{65}{96} \frac{D}{\xi^2_{\mathrm{D}}}
  \frac{\omega_{\mathrm{L}}}{\Omega_B^2} (2\omega_r + \omega_z).
\label{eq:tau_SD_harm}
\end{equation}
In our temperature range with ballistic quasiparticle transport $\tau_{\mathrm{SD}}^{-1} (T) \propto D(T) \propto \exp(-\Delta/k_{\mathrm{B}}T) \propto (\Delta f_{\mathrm{f}}(T) - \Delta f_{\mathrm{f}}^{\mathrm{i}})$, where $\Delta f_{\mathrm{f}}^{\mathrm{i}}$ is the intrinsic width of the fork. Thus the relaxation rate of the magnon condensate is as sensitive a thermometer as the damping of the oscillating object.

It is important for the interpretation of the measurements that the trapping frequencies have no essential temperature dependence; in our geometry the axial profile of the potential is fixed by the applied magnetic field and is constant at given $I_{\mathrm{min}}$, and the texture-determined radial potential is approximately temperature-independent at low temperatures $\lesssim 0.3T_{\mathrm{c}}$. However, the trapping frequencies have a dependence on magnetic field from two sources. First, the magnon mass scales as $m_{\mathrm{M}} \propto \omega_{\mathrm{L}} = \gamma H$. In a harmonic magnetic axial trap this results in a dependence $\omega_{z} \propto H^{-1/2}$. Second, in the radial direction $\omega_{r} (H)$ has a more complicated form since the spin-orbit energy in Eq.~(\ref{eq:spin-orbit}) includes both direct dependence on $H$ and indirect dependence of $\beta_l(r)$ distribution via the magnetic healing length $\xi_H \propto 1/H$~\cite{thune_hydros}. This leads to additional decrease of $\omega_r$ with increasing $H$. In the experiment $\omega_r$ and $\omega_z$ can be independently measured from the NMR spectra~\cite{magnon_relax_JLTP}.

\textit{Radiation damping} results from induction losses when the magnetization precesses and induces a voltage $V_{\mathrm{s}}$ across the pick-up coil. The induced voltage drives a dissipative current through the resistive impedance $R$ of the tank circuit, leading to Joule heating and the exponential relaxation of magnetization with rate $\tau_{\mathrm{RD}}^{-1} \propto V_{\mathrm{s}}^2/R$~\cite{magnon_relax_JLTP, raddamp}. In the $LC$ tank circuit which is used for pick-up in the present experiment, $V_{\mathrm{s}}$ has the standard Lorentzian resonance frequency dependence. This dependence is the main contribution to the variation of $\tau_{\mathrm{RD}}^{-1}$ with magnetic field. Since the magnetic flux through the pick-up coil, and thus the induced voltage $V_{\mathrm{s}}$, depends on the profile of the magnon density in the condensate, an additional weak $H$-dependence of $\tau_{\mathrm{RD}}^{-1}$ follows from the change of the spatial extent of the condensate as the trap shape changes with the field. The decrease of the condensate size reduces the induced $V_{\mathrm{s}}$ and the corresponding radiation damping.

Radiation damping increases with the increase of the quality factor $Q\propto 1/R$ of the tank circuit, so low-$Q$ circuitry is favored in magnon BEC relaxation measurements to minimize this effect. In the low temperature regime the radiation damping is independent of $T$ since neither the trapping potential nor the profile of the magnon density depend on $T$, see above.

Combining both relaxation mechanisms we get
\begin{equation}
\tau_M^{-1}(T,H) = A(2\omega_r(H) + \omega_z(H)) (\Delta f_{\mathrm{f}}(T) - \Delta f_{\mathrm{f}}^{\mathrm{i}}) + \tau_{\mathrm{RD}}^{-1}(H)~,
\label{eq:tau_M}
\end{equation}
where $A$ is a proportionality constant. Note that $A$ is approximately field-independent since in Eq.~(\ref{eq:tau_SD_harm}) $D \propto \omega_{\mathrm{L}}^{-1}$. This linear relation between $\tau_M$ and $\Delta f_{\mathrm{f}}$ is measured in Fig.~\ref{fig:t_calibration}a for two different magnetic fields and two different intrinsic fork widths $\Delta f_{\mathrm{f}}^{\mathrm{i}}$. The slopes of the fitted lines in Fig.~\ref{fig:t_calibration}a at the frequencies 623.2\,kHz and 832.6\,kHz are in the ratio $1.367:0.922 = 1.48$, while the ratio of $(2\omega_r + \omega_z)/2\pi$ in these measurements is $656\,{\mathrm{Hz}}: 453\,{\mathrm{Hz}} = 1.45$. Thus the slopes are described reasonably well by Eq.~(\ref{eq:tau_M}) but, as seen from the fits, this relation is valid only at the lowest temperatures. The actual relaxation rates start to deviate at higher $T$ as the approximate exponential temperature dependence of $D$ ceases to be valid.

The temperature-independent constant on the RHS of Eq.~(\ref{eq:tau_M}) includes both the intrinsic fork width and the radiation damping of the magnon condensate. The latter can be extracted from a measurement of $\tau_M^{-1}$ as a function of $H$ at constant $T$, Fig.~\ref{fig:t_calibration}b. If the pick-up tank circuit has a quality factor $Q\gg1$ then the radiation damping term on the RHS of Eq.~(\ref{eq:tau_M}) varies with $H$ much faster than the spin-diffusion term, allowing for convenient fitting of the radiation damping to the square of a Lorentzian response with a known $Q$. In particular, for measurements performed at the frequency of the tank circuit resonance, like those in Fig.~\ref{fig:t_calibration}a, the magnitude of the radiation damping is the height of the peak in Fig.~\ref{fig:t_calibration}b above the spin-diffusion background. To overcome the problems in this measurement due to decreased sensitivity away from the resonance of the $LC$ tank circuit we increased the amount of averaging of the signals as we moved further away from the resonance.

At $T=0$ we have $\Delta f_{\mathrm{f}}(T) = \Delta f_{\mathrm{f}}^{\mathrm{i}}$ and $\tau_M^{-1} = \tau_{\mathrm{RD}}^{-1}$, as seen from Eq.~(\ref{eq:tau_M}). Therefore by extrapolating the linear dependence in Fig.~\ref{fig:t_calibration}a until $\tau_M^{-1}$ reaches the measured value of the radiation damping one can find the intrinsic damping of the thermometer fork. This procedure is illustrated in Fig.~\ref{fig:t_calibration}c. Note that in the two measurements (marked with squares and triangles), which were performed at different frequencies in the same cooldown, both the slopes of the linear dependences and the radiation damping values are different. However, the resulting values of $\Delta f_{\mathrm{f}}^{\mathrm{i}}$ are found to be identical. In the measurement a few months later (filled circles) the slope and the radiation damping were found to be very close to the earlier measurement at the same frequency. The fork, however, had deteriorated during the intervening period and this is clearly detectable as an increase in intrinsic damping.

If the thermal contribution to the damping of a mechanical-oscillator-based thermometer is known, then the transfer of this calibration to the relaxation rate of the magnon condensate requires a measurement of the dependence $\tau_M^{-1}(\Delta f_{\mathrm{f}})$ and of
$\tau_{\mathrm{RD}}^{-1}$, as described above, which simultaneously provides a reliable value for $\Delta f_{\mathrm{f}}^{\mathrm{i}}$. If such a calibration is not available, the magnon condensate can still be used as a stand-alone thermometer if $\omega_r$ and $\omega_z$ are additionally measured and the theoretical value of the diffusion coefficient $D$ from Ref.~\cite{Einzel_JLTP84} is used. As for any resonator, the usability of magnon BEC based thermometry is limited by its ``intrinsic'' damping, which in this case is the level of radiation damping $\tau_{\mathrm{RD}}^{-1}$. At low enough temperatures $\tau_{\mathrm{SD}}^{-1} \ll \tau_{\mathrm{RD}}^{-1}$, preventing precise temperature measurements. In the measurements presented here this limit is not reached and at the lowest measured $T$ we have $\tau_{\mathrm{SD}}^{-1} \gtrsim \tau_{\mathrm{RD}}^{-1}$.

To test a standalone usage of the magnon BEC as a thermometer, we performed direct calculations of $\tau_{\mathrm{SD}}$ for conditions of Fig.~\ref{fig:t_calibration}, starting from Eq.~\eqref{eq:spinwave_schrode}, as described above. The magnetic field profile in the trap is determined from the spectrum of axial excited states and from the NMR line shape in normal $^3$He~\cite{magnon_relax_JLTP}. The textural part of the potential is calculated as described in Refs.~\cite{thune_hydros, kopu_JLTP146} with the value of the Leggett frequency $\Omega_{\mathrm{B}}$ based on Ref.~\cite{thune_hydros}. The value for $\xi_D$ is found from Ref.~\cite{hakonen_JLTP76} and the theoretical value of $D$ from Ref.~\cite{Einzel_JLTP84} is used. Experimentally measured value of $\tau_{\mathrm{RD}}$ is taken into account. Using these calculations we have converted relaxation measurement in Fig.~\ref{fig:t_calibration} to temperature and found that the difference with the temperature determined from the fork is less than $5\cdot10^{-3} T/T_{\mathrm{c}}$ in the temperature range covered. (In terms of $A$ factor in Eq.~(\ref{eq:tau_M}) this means about 25\% maximum difference between the fit in Fig.~\ref{fig:t_calibration} and calculations.) We consider this agreement to be reasonably good given that in our setup there is a substantial separation between the condensate and the thermometer fork, and thus some temperature difference may develop between them owing to heat leaks into the container and diffusive scattering of thermal quasiparticles from the container walls. Moreover, the fork calibration itself might have a comparable uncertainty.

The calibration measurements described so far have been done with the container filled with B phase in low or zero magnetic field. In the following section we will be interested in the small temperature increase in the top part of the sample cylinder caused by increasing the magnetic field or by creating an A-phase layer in the middle of the sample. These relative temperature measurements were conducted in the linear $\tau_M^{-1}$ vs $\Delta f_{\mathrm{f}}$ regime so the error in determining the constant temperature offset should be unimportant.

\begin{figure}
\centerline{\includegraphics[width=\textwidth]{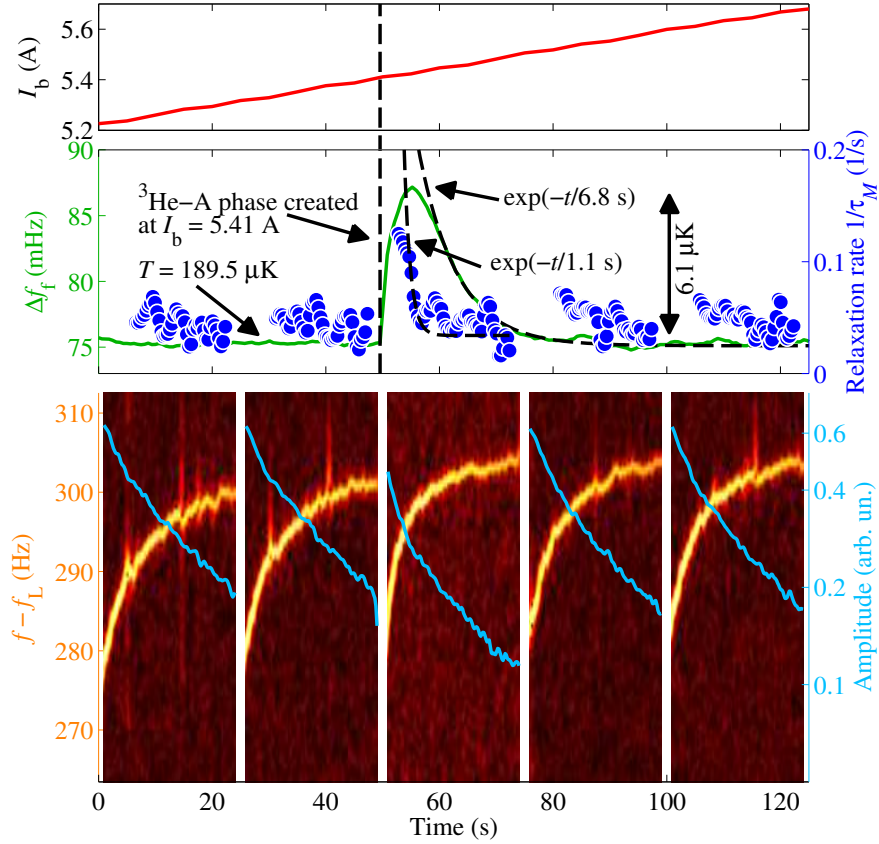}}
\caption{\label{fig:A_formation} (Color online) Thermal response from the formation of $^3$He-A in overmagnetized $^3$He-B. The topmost panel shows the current $I_{\mathrm{b}}$ in the barrier magnet, which is slowly increased while the whole sample is originally filled with B phase. The A phase forms at current $I_{\mathrm{b}} = 5.4\,$A, well above the equilibrium value $I_{\mathrm{b}} = 4.8\,$A, and quickly expands to form a layer approximately 0.7\,cm thick, see Fig.~\ref{fig:meas_setup}. In this non-equilibrium process heat is released. This is seen in the middle panel where the fork width $\Delta f_{\mathrm{f}}$ (solid line) and the relaxation rate of the magnon condensate $\tau_M^{-1}$ (circles) are plotted as a function of time. Note the faster response of the condensate compared to the fork. The bottom panel displays the relaxing magnon BEC signals after five consecutive excitation pulses at 25\,s intervals. The yellow-red colors show color-coded time-dependent Fourier spectra of the signal from the pick-up coil with the peak (bright yellow) corresponding to the coherent precession of the condensate. The frequency of precession increases during the relaxation owing to the self-modification of $\omega_r$ by the condensate~\cite{self-trapping_PRL}. The blue lines show the time dependence of the total amplitude of the precession signal (logarithmic scale) from which $\tau_M$ in the middle panel is extracted using a 3\,s window for each point with a 0.3\,s step between the points.}
\end{figure}

\section{Thermal response from the creation of A-phase}
\label{sec:therm_resp}

Above we discussed temperature measurements in the steady-state situation. Here we compare the response rates of the fork and of the magnon condensate to the heat pulse generated by a non-equilibrium B$\rightarrow$A transition. If the transition from the B to the A phase occurs close to thermodynamic equilibrium, the latent heat of the transition is absorbed and in a thermally isolated volume the corresponding cooling can be seen~\cite{AB-cooling_lancaster}. When the transition is triggered by the increase of the magnetic field, it is often observed that the B phase remains in a metastable overmagnetized state above the equilibrium field of the transition $H_{\mathrm{AB}}$, and the transition finally happens in a field $H_{\mathrm{t}} > H_{\mathrm{AB}}$. In the case of such non-equilibrium A-phase formation a relatively large volume is converted quickly to A phase. This process supposedly involves rapid propagation of the AB interfaces separating the growing A-phase layer in the B-phase column. The motion of the AB interface has an associated friction~\cite{AB-friction_lancaster,AB-friction_YIP,AB-friction_kopnin,Lancaster-microkelvin}. When a sufficiently wide layer of A phase is created, then the net thermal effect becomes positive and a heating spike is observed. On repetition of the field increase across the transition some spread and history dependence of $H_{\mathrm{t}}$ is observed~\cite{AB-history_PRL85}.

An example of such a measurement is shown in Fig.~\ref{fig:A_formation}. Here the current $I_{\mathrm{b}}$ in the barrier magnet is swept upwards until the A phase is formed. The moment of transition is seen as a sudden rise in $\Delta f_{\mathrm{f}}$. At the same
time the relaxation rate of the magnon BEC gets faster. During the measurement the condensate is refilled by periodic excitation pulses at 25\,s intervals after which the decay rate of the precession signal is measured. The relaxation rate $1/\tau_M$ is defined from the relaxation signals using a 3\,s sliding window. We exclude parts of the signal at large frequency shifts (and thus large magnon densities) where other relaxation effects beyond those described in this paper become important.

As shown by the fitted exponential decays (broken lines) in Fig.~\ref{fig:A_formation}, the rate of recovery to the original level is much faster for the condensate (time constant 1.1\,s) than for the fork (6.8\,s). Both of these systems are high Q resonators and thus have slow response rates at low temperatures. The reason for the difference we see is related to our measuring techniques. The fork is measured in the traditional ``continuous wave'' mode, where the excitation is kept constant and the amplitude of the response is recorded. In this case the response rate is limited by the Q value as $1/(\pi \Delta f_{\mathrm{f}}) = 4.2\,$s. The magnon BEC is measured in the pulsed mode, and ``instantaneous'' decay rate is found from the derivative of the amplitude of the decaying signal. In this case the time resolution is limited by the signal to noise ratio in the amplitude measurement. In principle, the fork could also be measured in the pulsed mode. However, to restart oscillations after the decay, one should apply an excitation pulse to the fork which should be the stronger the shorter the dead time is required to be. In contrast, the magnon BEC can be refilled via off-resonant excitation in a short time below 0.1\,s independent of the strength of the excitation pulse.

\begin{figure}
\centerline{\includegraphics[width=0.7\textwidth]{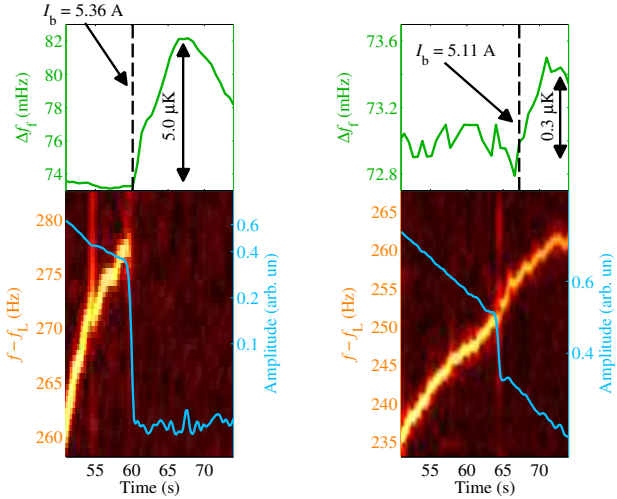}}
\caption{\label{fig:A_formation_extra} (Color online) Two examples of the B to A transition with smaller overmagnetization of the B phase than in Fig.~\ref{fig:A_formation}. The larger the overmagnetization the higher the temperature increase is during the A-phase formation process. \textit{(Left)} Magnon BEC rapidly collapses due to the large sudden increase in temperature or due to dynamics related to the B$\rightarrow$A transition. \textit{(Right)} A small increase in temperature is seen only as a slight kink in the decaying signal.}
\end{figure}

Two more examples of A-phase creation with smaller overmagnetization are presented in Fig.~\ref{fig:A_formation_extra}. For the transition at $I_{\mathrm{b}} = 5.36\,$A the temperature increase as seen by the fork is larger than that for the transition at $I_{\mathrm{b}} = 5.11\,$A, as expected. However, the rapid collapse of the magnon BEC in the left plot cannot be explained by the modest temperature increase alone as seen by the fork. It seems that the dynamics involved in the formation of the A-phase layer is more violent than can be resolved from the slow fork response. In the plot on the right a much smaller layer of A phase is formed and the condensate survives the transition. In this case an interesting new feature is the apparent delay of the fork signal compared to the condensate response. This feature also remains unexplained.

One should consider the observations in Figs.~\ref{fig:A_formation} and \ref{fig:A_formation_extra} as a proof of concept. The current setup of the magnon BEC experiment is not optimized to be used as a thermometer. For instance the ``dead time'' between separate relaxation traces in Fig.~\ref{fig:A_formation} is too long. This is caused by the time required to transfer the NMR trace to the acquisition computer and by over-exciting the magnon BEC into the non-linear relaxation regime. In principle it will not be difficult to reduce the dead time to the 0.1\,s needed for the thermalization of pumped magnons. An even better approach might be to replenish the condensate continuously by pumping to the excited state of the trap in a feedback loop so that the amplitude and frequency of the ground-state precession remain constant. In this case the amplitude of the feedback pumping would provide a continuous reading of the condensate relaxation rate and thus of the temperature.

\section{Measurements of quasiparticle impedance of the $^3$He-A layer}
\label{sec:T-diff}

In Lancaster thermal resistance was studied in a long column filled with superfluid $^3$He at temperatures $150-200\,\mu$K and at zero pressure~\cite{AB-brane_JLTP}. The thermal resistance over 50\,mm of column length was measured in three different configurations: (1) the sample cell was filled with B phase in low magnetic field; (2) part of the sample was magnetic-field-distorted B phase just below $H_{\mathrm{AB}}$; and (3) an A-phase layer was created between two B-phase volumes. Similar configurations can be created in our setup. In the Lancaster measurements it was found that the thermal resistance with the A-phase layer is almost doubled when compared to the B-phase resistance in low field. This result was the same for a wide range of thicknesses of the A phase, indicating that the extra resistance is associated with AB interfaces rather than with bulk A phase. It was suggested that Andreev reflection of quasiparticles with energies below the maximum gap at the interface is the source of the extra resistance. Since the difference between the maximum and minimum gap values remains constant at low temperatures, while the population of quasiparticles with energies above the minimum gap rapidly decreases when $T\rightarrow0$, then this Andreev thermal resistance of the interface should rapidly increase with decreasing temperature. This qualitative conclusion can be checked in our setup.

\begin{figure}
\centerline{\includegraphics[width=\textwidth]{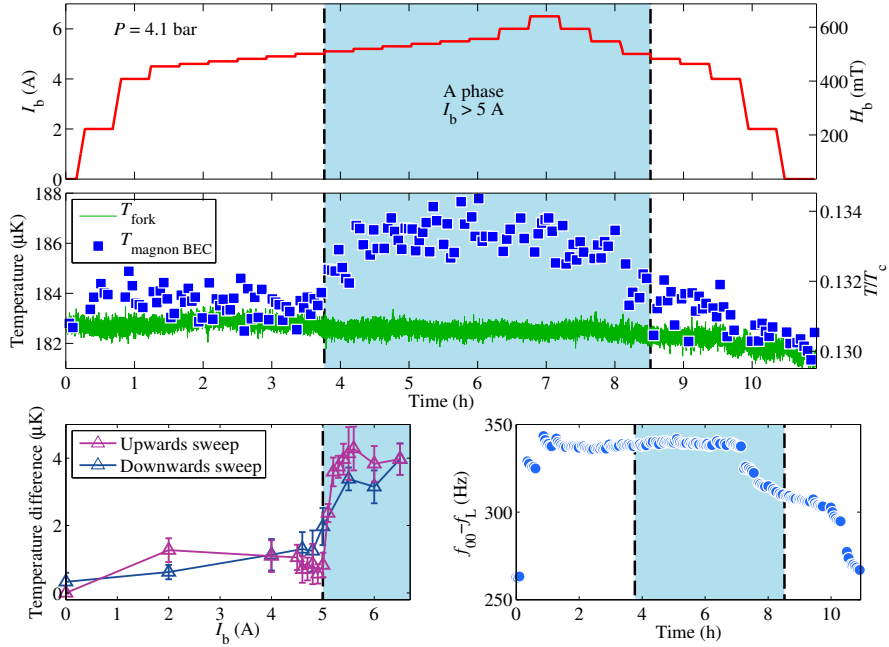}}
\caption{\label{fig:A-phase_T_diff} (Color online) Thermal resistance effect from the $^3$He-A layer in the sample cylinder. \textit{(Top)} A sweep of the current in the barrier magnet $I_{\mathrm{b}}$ as a function of time. The A phase exists in the middle of the sweep. \textit{(Middle)} Temperature at the bottom of the cell as measured with the quartz tuning fork and at the top of the cell as determined from the relaxation rate of the magnon BEC. Calibration from Fig.~\ref{fig:t_calibration} is used (which makes temperature readings of the two thermometers coincide at $t=0$). \textit{(Lower left)} The difference $\Delta T$ between the two temperature records of the middle panel plotted against $I_{\mathrm{b}}$. The average difference when the A phase exists in the sample ($I_{\mathrm{b}} > 5\,$A) is $\Delta T =4.0\pm0.3\,\mu$K. With the magnetized B-phase layer $\Delta T = 0.8\pm 0.2\,\mu$K ($I_{\mathrm{b}}=4\div5\,$A). Within error limits, $\Delta T$ after the A-phase annihilation is the same as before the A-phase creation. \textit{(Lower right)} Precession frequency of the magnon condensate during the same measurement.}
\end{figure}

Since the main purpose of our setup is to study magnon condensation, we do not have a heater at the closed end of the sample tube to control the heat flux along the tube at will. Instead the heat flux is set by the background heat leak. Thus we cannot measure absolute values of the thermal resistance. From the temperature change at the top of the tube for a given flux we can only deduce changes in the thermal resistance. We thus take the configuration with the whole tube filled with low-field B phase as a base (``zero'') value of thermal resistance. Compared to this base value we find an increased resistance with distorted B phase in high magnetic field and an increased resistance in the presence of the AB interfaces. An example of such a measurement is presented in Fig.~\ref{fig:A-phase_T_diff}. Here the current in the barrier magnet $I_{\mathrm{b}}$ is first slowly increased in steps. At each stable value the relaxation rate of the magnon condensate $\tau_M^{-1}$ is measured and converted to temperature using the calibration from Fig.~\ref{fig:t_calibration} as explained in the previous section. The A phase is created in the middle of the sample column at $I_{\mathrm{b}}=5.1\,$A. The temperature at the top of the sample is seen to increase when the A phase is present. With decreasing $I_{\mathrm{b}}$ the behavior is reversed and the temperature at the top returns to its original value. The temperature at the bottom, determined with the tuning fork, remains constant during the measurement, indicating that the heat leak is also constant. This is important for the interpretation of the measurement. We conducted this kind of measurement multiple times and found the results were reproducible.

For our analysis we assume that the 12\,pW heat leak into the sample is evenly distributed along the sample tube. For $I_{\mathrm{b}}=6$\,A (see Fig.~\ref{fig:meas_setup}) this gives 4\,pW for the heat leak above the A-phase layer, 1\,pW into the A phase, and 7\,pW below the A phase. As seen in Fig.~\ref{fig:A-phase_T_diff} the temperature increase in the top when the B phase in the middle of the sample is magnetized close to $H_{\mathrm{AB}}$ is $\Delta T = 0.8\,\mu$K, and when the middle section is in the A phase $\Delta T = 4\,\mu$K (which also includes the contributions from the magnetized B phase close to the AB interfaces).

Most of the thermal resistance between the two ends of the sample, when the A-phase layer is present in between, comes from the ballistic quasiparticles in the top being prevented from crossing the first B-A boundary since they do not have energy higher than the highest energy gap in the boundary region. The distorted B-phase texture and Zeeman splitting of excitation energies close to the top interface increases the resistance further~\cite{AB-brane_JLTP}. This allows us to assume that the $4\,\mu$K temperature increase is caused by a 4\,pW heat leak above the A phase alone, so we can calculate an extra thermal resistance from the AB boundary region of about $1\,\mu$K/pW. When we take into account the area of our AB interface and convert the result to boundary ``resistivity'', we get $\rho_{\mathrm{A+B}} \approx 30\,\mu$K/pW mm$^2$. This value of $\rho_{\mathrm{A+B}}$, measured here at $T = 0.131T_{\mathrm{c}}$, is larger than $\rho_{\mathrm{A+B}}=8.5\,\mu$K/pW mm$^2$ found in Ref.~\cite{martin_thesis} at $T = 0.15T_{\mathrm{c}}$, as might be qualitatively expected. To compare these results more quantitatively we use a simple one-dimensional model of quasiparticle propagation and additionally ignore the backward flux of ballistic quasiparticles. This gives a scaling for the thermal resistance,
\begin{equation}
R \propto \frac{\exp(\Delta_{\mathrm{m}}/k_{\mathrm{B}} T)}{\Delta_{\mathrm{m}}+k_{\mathrm{B}} T}~.
\label{eq:R}
\end{equation}
Here $\Delta_{\mathrm{m}}$ is the largest energy gap encountered by the quasiparticles. The largest gap is about 1.15 times the BCS gap for the A phase and 1.2 times the largest gap for Zeeman-split quasiparticles in the B phase. Using Eq.~(\ref{eq:R}) to extrapolate the measurements from Ref.~\cite{martin_thesis} to $T = 0.131T_{\mathrm{c}}$ we get $\rho_{\mathrm{A+B}} \sim 60\,\mu$K/pW mm$^2$. This is in satisfactory agreement with our measured value, considering the rapid dependence of $\rho_{\mathrm{A+B}}$ on the temperature of the measurement. (A more sophisticated thermal model can be found in Ref.~\cite{martin_thesis}.)

\textit{Modification of the magnon trap.} The magneto-textural trap for magnons is sensitive to changes in both the textural and the magnetic confinement energies, which affect the trapping frequencies $\omega_r$ and $\omega_z$. These in turn control the precession frequency $\omega_{00} = \omega_{\mathrm{L}} + \omega_r +\omega_z/2$ of the ground-state condensate. In our setup the solenoid which creates the barrier field is designed to provide minimum interference to the NMR spectrometers, although the field compensation is not perfect. It is expected that at some level the residual field of the barrier solenoid at the position of the NMR spectrometer changes $\omega_{\mathrm{L}}$ while the increased field inhomogeneity affects $\omega_r$ and $\omega_z$. One would expect that these contributions are directly controlled by $I_{\mathrm{b}}$ and are history-independent. When $I_{\mathrm{b}}$ is changed in the measurements of Fig.~\ref{fig:A-phase_T_diff}, a change in $f_{00}=\omega_{00}/2\pi$ is indeed observed (plot on the lower right). However, it is also immediately seen that the behavior is history-dependent: $f_{00}$ shows a sudden increase when $I_{\mathrm{b}}$ starts to increase, then it remains at a stable value until the maximum $I_{\mathrm{b}}$ is reached, without any change when A phase appears, and finally displays a gradual decrease towards the original value as $I_{\mathrm{b}}$ is swept down. This in principle can be caused by pinning of the magnetic flux in the superconducting parts of the setup, leading to hysteretic behaviour of the magnetic field profile and thus of $f_{00}$. Alternatively, the order-parameter orbital texture can be involved. At present very little is known about the textural dynamics in ultra-cold $^3$He-B, but in the following section some examples of non-trivial textural dynamics are presented.

\section{Textural response to creation and annihilation of $^3$He-A}

An interesting conclusion from earlier measurements on the thermal resistance of the AB interface is the suggestion that order-parameter defects are left behind in the B-phase column after the removal of the A-phase layer~\cite{AB-brane}. The signature of the defects is an additional resistance in the quasiparticle transport in B phase after A-phase annihilation when compared to that measured before the A-phase layer was created. Annihilation of the A-phase layer proceeds via the collision of an AB interface with a BA interface. It has been speculated that this process might be similar to brane annihilation in cosmology, where it has been suggested that colliding branes and anti-branes leave behind topological defects such as cosmic strings~\cite{brane-inflation_PRD}.

\begin{figure}
\centering
\flushleft
\includegraphics[width=0.99\textwidth]{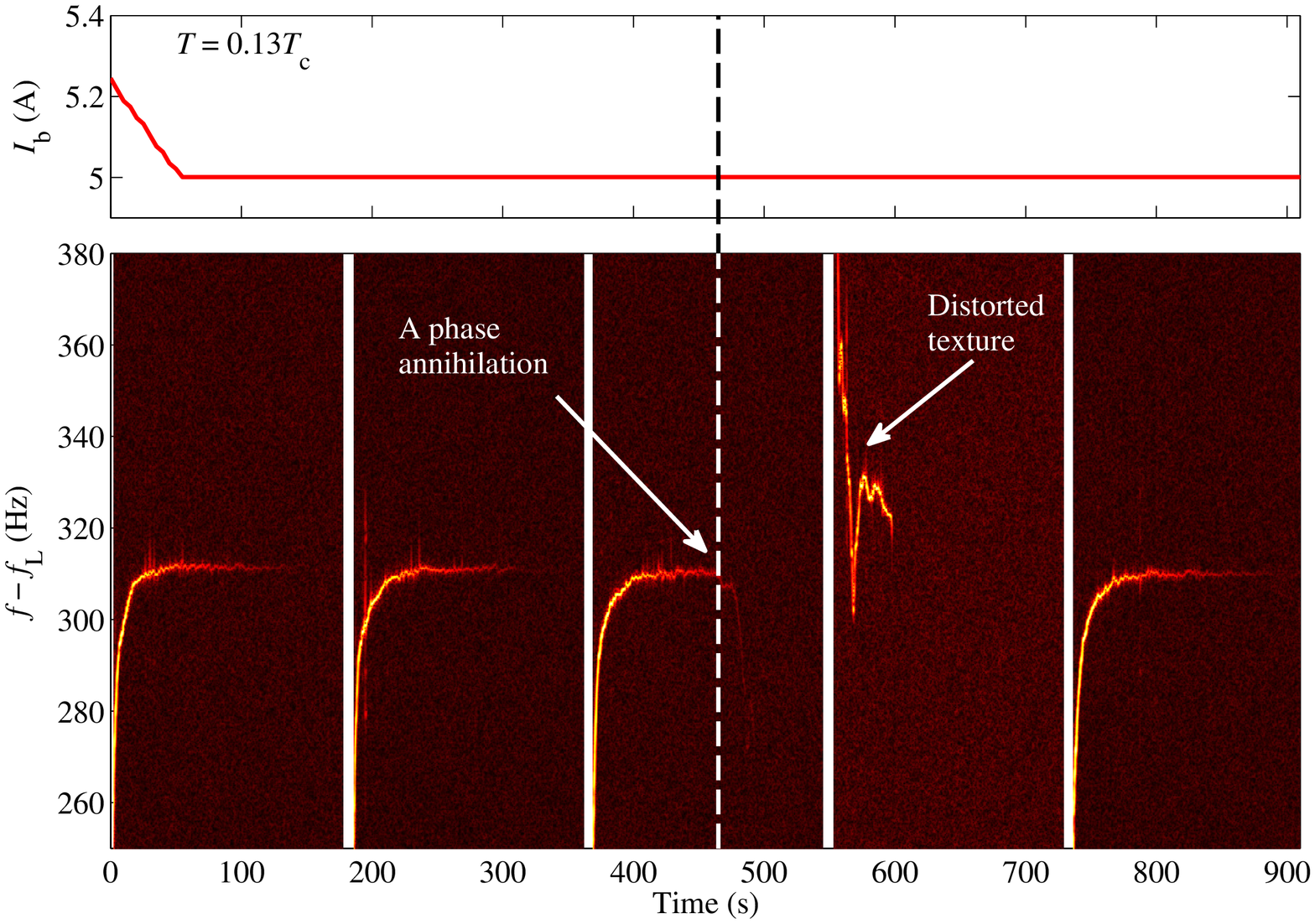}
\hspace{-3.1cm}
\llap{\raisebox{6.6cm}{
\includegraphics[width=0.4\textwidth]{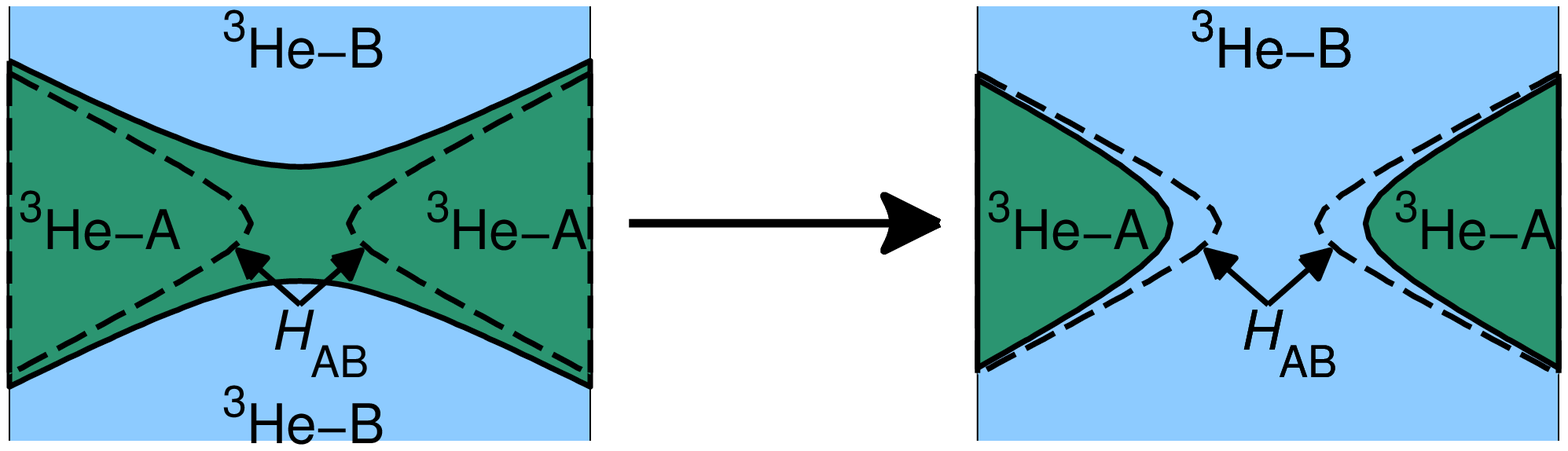}
}}
\caption{\label{fig:A-phase_annih} (Color online) Distortion of the orbital texture of the order parameter in $^3$He-B by rapid annihilation of an A-phase layer. The current $I_{\mathrm{b}}$ in the barrier magnet is swept down from above and left constant at the value where the AB interface profile at $H=H_{\mathrm{AB}}$ has a toroidal shape with a central hole. The already existing A-phase layer is left in a metastable state without the hole, owing to the surface tension of the AB interface (top panel). At some later time, as a result of some disturbance, the interface relaxes to its equilibrium shape with a hole. The rapid non-equilibrium motion of the AB interfaces during this highly irreversible event causes disturbances in the orbital texture in the B-phase column, which at time $t = 460\,$s reach the volume of the magnon condensate. These disturbances are seen as fluctuations of the precession frequency of the magnon BEC as demonstrated by the time-dependent Fourier spectra of the NMR signal shown in the bottom row. The condensate has been refilled with magnons a few times during the measurement (white vertical gaps). After about 200\,s the texture relaxes to its original state.}
\end{figure}

The measured value of the additional resistance was not reproducible but varied between measurements, one would assume dependent on the configuration of the defects created. It has been measured to be as high as 50\% of the resistance difference between the magnetized B-phase resistance and the AB interface resistance~\cite{AB-brane}, which would correspond to an extra temperature difference $\Delta T = 1.6\,\mu$K in our setup after the A-phase annihilation, Fig.~\ref{fig:A-phase_T_diff}. However, as explained in the previous section, in the state after the annihilation of the A-phase layer we have not observed, within our sensitivity, any extra thermal resistance which could be attributed to order-parameter defects as suggested in Ref.~\cite{AB-brane}. Nevertheless, we have observed another, non-thermal, signature based on the direct influence of the order-parameter texture on the magnon condensate precession. An example is provided in Fig.~\ref{fig:A-phase_annih}. Here the measurement starts at $I_{\mathrm{b}} = 6.5\,$A with an A-phase layer in the sample. Then $I_{\mathrm{b}}$ is swept down in steps of about 0.3\,A and the waiting time at each current is 30 minutes. We periodically excite the magnon BEC with excitation pulses of constant amplitude and duration and monitor the frequency of the precession using the Fourier transform of the pick-up signal. Standard decay curves with the frequency relaxing exponentially upwards are usually observed, but five consecutive decay signals in Fig.~\ref{fig:A-phase_annih} capture a special event. Here the A phase relaxes from a metastable configuration of a thick layer across the cylinder to a toroidal configuration with a central hole~\cite{ABhole}. This is exactly the process leading to ``brane annihilation'' and studied in Ref.~\cite{AB-brane}. Irregular fluctuations of the condensate precession frequency are seen after this event and are caused by fluctuations of the orbital texture modifying $\omega_r$. We have not seen fluctuations reach the volume of the condensate in all A-phase annihilation events; sometimes no special feature is observed at all, and sometimes the condensate simply collapses, probably due to the thermal signal related to dissipation from rapid movement of the frictional AB interface.

It is known that the existence of the AB interface always affects the B-phase texture since the orbital anisotropy axis in the B phase is pulled parallel to the interface~\cite{thune_AB}. Thus any disturbance or movement of the AB interface must generate a disturbance of the B-phase texture. However, the time scale of the observed fluctuations much exceeds the fraction of a second which the reconfiguration of the A-phase layer takes. Thus it is tempting to ascribe them to textural waves emitted by a B-phase defect left after the annihilation of the AB interfaces over a large part of the cylinder's cross-section, as proposed by the Lancaster groups. Another non-thermal signature of an order-parameter defect, possibly created by annihilation of the A phase, was seen in the measurements of the Josephson effect in $^3$He-B \cite{joseph-wall}. The extended time scale of the fluctuations may reflect the time needed by the defect to settle into a stable configuration or disappear. For example, remanent vortices can move in the long sample cylinder for hours even in the absence of applied flow at low temperatures~\cite{dynremnant}, and those are known to modify the trapping potential for magnons via their contribution to the textural energy~\cite{eltsov_JLTP162}.

\begin{figure}
\centerline{\includegraphics[width=\textwidth]{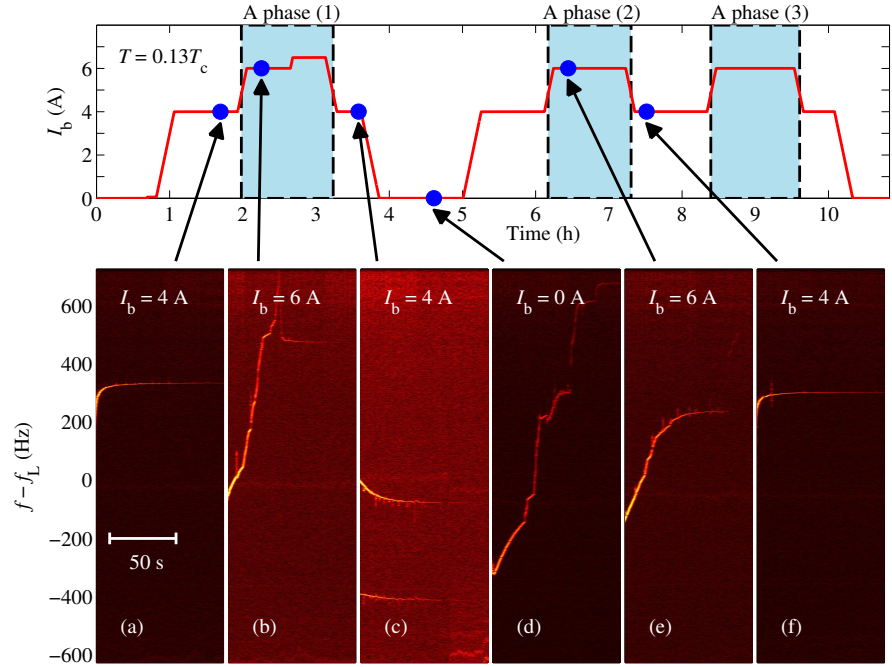}}
\caption{\label{fig:A-phase_defect} (Color online) Long-lasting B-phase texture fluctuations triggered by the formation of A phase from strongly overmagnetized B phase. The top panel shows the current $I_{\mathrm{b}}$ as a function of time while in the bottom panel examples of signals of the condensate precession measured in the same way as in Fig.~\ref{fig:A-phase_annih} are shown. The fluctuating signal appears when the A phase is created for the first time and disappears after the A phase is annihilated for the second time. The third cycle of creation and annihilation of the A phase does not cause any notable effect on the frequency of the condensate precession beyond that shown in Fig.~\ref{fig:A-phase_T_diff}.}
\end{figure}

One should be careful, though, in attributing the formation of defects solely to annihilation of AB interfaces. In the event in Fig.~\ref{fig:A-phase_annih} another important feature is the fast motion of the interface from the original to the final configuration. One might ask which process is more important for the creation of the textural signal, annihilation or fast motion? Insight into this question is found from Fig.~\ref{fig:A-phase_defect}. Here multiple relatively fast sweeps of $I_{\mathrm{b}}$ with large steps have been performed. During the fast upsweep of the magnetic field it is difficult to identify the exact moment when the A-phase layer is formed, but from the heating signal seen by the fork one can conclude that the overmagnetization of the B phase was in this case substantial. The magnon condensate was refilled every three minutes over the whole measurement time of ten hours. In the bottom row of Fig.~\ref{fig:A-phase_defect} six example signals from different moments of the run are shown. All the signals before the first A-phase creation and after the second A-phase annihilation showed no peculiar textural dynamics, as the example signals (a) and (f) demonstrate. All the relaxation signals recorded from the first A-phase creation until the second A-phase annihilation show fluctuating textural behavior. Four example signals from this sequence are in the middle of the bottom row in Fig.~\ref{fig:A-phase_defect}. As can be seen, and in contrast to Fig.~\ref{fig:A-phase_annih}, here the textural fluctuations last for 5 hours! Moreover, they persist even when there is no A phase present in the sample (example signals (c) and (d)) and even when the current $I_{\mathrm{b}}$ is reduced to zero (example signal (d)). These very long-living fluctuations have also been seen on other occasions where the current $I_{\mathrm{b}}$ was swept relatively quickly, but not in all such cases. It is difficult to imagine any other source of such long-lasting textural fluctuations except a relatively stable order-parameter defect which would be moving in our long sample tube. Textural fluctuations lasting for hours in $^3$He-B have been previously seen in connection with a spin-mass vortex at much higher temperature~\cite{SMV}.

Note, however, that the fluctuating texture above is formed during the \textit{creation} of the A-phase layer from overmagnetized B phase. In this process the AB interfaces move rapidly. In one of the models of the supercooled A$\rightarrow$B transition the transition is considered to proceed through coalescence of independently forming B-phase blobs~\cite{Bunkov-AdV}. One might think about applying this scenario to the overmagnetized B$\rightarrow$A transition where individual blobs of A phase form and ultimately grow to coalesce to form the layer. In this scenario ``brane annihilation'' would also occur during this rapid A-phase formation process, but in the opposite direction (B phase disappears), and the remnants, if any, would be buried inside the A-phase layer. In this situation it is not clear how such remnants would affect the behavior of the texture on the B-phase side. On the other hand, it is known that a moving AB interface can lead to the creation of defects in the B phase, both spin-mass~\cite{SMV-AB} and regular~\cite{AB-instability} vortices.

The formation of a B-phase defect during A-phase creation does not contradict the earlier thermal resistance measurements in Lancaster: when the A phase is present, only the total resistance of the A and B phases is monitored in those measurements, so a defect could already exist then, as well as after the A-phase annihilation.

The fluctuations in the B-phase texture, which modulate the magnon condensate trapping potential, are seen to take rather diverse appearances. We observe fast changes and jumps of the precession frequency both upwards and downwards. Sometimes the frequency drops below the bottom of the original trap (marked with zero frequency shift in Fig.~\ref{fig:A-phase_defect}), which means that the position of the trap changes to the region of the smaller magnetic field. Sometimes simultaneous signals from several condensates are observed as in the example signal (c) in the bottom row of Fig.~\ref{fig:A-phase_defect}, which indicates splitting of the trap to several spatially isolated energy minima. The magnon BEC relaxation, nevertheless, in all these examples is still exponential yielding no reason to believe that additional relaxation processes to those discussed above are present. The relaxation rate, though, is often increased in the fluctuating state. This is understandable since the gradients of the magnon density in the strongly distorted traps would be larger than in the original trap and thus the rate of spin diffusion would increase (see Eq.~(\ref{eq:tau_SD})). This also means that using the relaxation rate of magnon BEC as a reliable measurement of temperature is impossible when defects modify its potential trap since a well-defined condensate shape is crucial for thermometry.

\section{Conclusions}
We have described a new approach to temperature measurement in ultra-cold superfluid $^3$He-B. It is based on the relaxation processes in trapped Bose-Einstein condensates of magnon quasiparticles. The temperature-sensitive relaxation process used for the measurements is the spin transport through the normal component in the ballistic regime. The method is contactless, potentially self-calibrating, and fast.

We have applied this technique to measure the thermal resistance of the AB interface in the ballistic regime of quasiparticle propagation and obtained a result which is in reasonable agreement with earlier measurements performed in Lancaster. We could not confirm their conclusion about the increased thermal resistance of the B phase after the annihilation of AB and BA interfaces. However, we have found a non-thermal signature after this event in the form of long-living fluctuations of the orbital B-phase texture. Such non-stationary dynamics lasting minutes and even hours is usually associated with the motion of relatively stable order-parameter defects, like vortices. We observe the formation of such defects both during the annihilation and the creation of the A-phase layer,
indicating that the formation of the B-phase defects is not related to brane annihilation alone, but may also be triggered by a rapid motion of the AB interface. The definite picture of the defect formation, however, should be supplied by further research.

\begin{acknowledgements}
We thank M. Krusius and P. Skyba for stimulating discussions. This work has been supported in part by the EU 7th Framework Programme
(FP7/2007-2013, Grant No. 228464 Microkelvin) and by the Academy of Finland through its LTQ CoE grant (project no. 250280). The research was done using facilities of the Cryohall infrastructure supported by Aalto University and Academy of Finland. P.J.H. acknowledges financial support from the V\"{a}is\"{a}l\"{a} Foundation of the Finnish Academy of Science and Letters.
\end{acknowledgements}

\section*{Appendix: Calculation of the spin diffusion coefficient}

The expression for the transverse spin diffusion tensor $D^\perp_{ij}$  \cite{Einzel_JLTP84,mark-mukh} is rather complicated and its correct numerical
evaluation provides some challenge. Here we describe one recipe for
successful calculation of the spin diffusion and also provide a
low-temperature approximation from which the origin of the exponential
dependence of $D$ on temperature becomes clear. We discuss only transverse
spin diffusion and thus omit the superscript `$\perp$' from the following equations.

The theoretical expression for $D$ is
\begin{equation}
D_{ij}(T, \ol) =
  \frac{\vF^2}{\chi'}\ \tp
  \  \int \frac{\sm - i(\tp\op)\sp}{1+(\tp\op)^2\sp}
  \  k_i k_j\ \phi_k\ d\xi_k\frac{d\Omega_k}{4\pi}.
\label{eq:D_Einz}
\end{equation}
Here $\vF$ is the Fermi velocity, $\chi'$ is the ratio of susceptibilities
of superfluid helium and ideal Fermi gas $\chi'=\chi_{\rm B}/\chi_{\rm N0}$, $\mathbf{k}$ is the wave vector and $\xi_k$ is the
energy of the quasiparticles. The integration is performed over the energy
and direction of the wave vector. We also make use of the following notation
$$
\op = \ol+\oe,\quad
\tp = \frac{\tdp}{1-i\ol\tdp},
$$
$$
\sp = u^2 + k_\parallel^2(1-u^2),\quad
\sm = 1-\frac{k_\perp^2}{2}(1-u^2),
$$
$$
u=\frac{\xi}{\sqrt{\xi^2+\Delta^2}}, \quad
\phi_k = \frac{1}{2T}\left(\cosh\frac{\sqrt{\xi^2+\Delta^2}}{2T}\right)^{-2},
$$
$$
k_\perp = \sqrt{k_x^2+k_y^2} = \sqrt{1-\kpa^2}, \quad
k_\parallel = k_z,
$$
where $\tau$ is the quasiparticle relaxation time and $\oe = \lambda \ol$ is
the Landau molecular field.

The integration over angles in Eq.~(\ref{eq:D_Einz}) can be performed
analytically, resulting in the following integrals over the energy which can be
evaluated numerically,
\begin{equation}
D_{xx} =
  \frac{v_F^2}{\chi'}\ \tp
  \ \int_{-\infty}^{\infty}
  \ \frac12 (I_1-I_2)\ \phi_k\ d\xi_k,
\qquad
D_{zz} =
  \frac{v_F^2}{\chi'}\ \tp
  \ \int_{-\infty}^{\infty}
  \ I_2\ \phi_k\ d\xi_k.
\label{eq:D_i}
\end{equation}
Here
$$
I_1 = \int_0^1 \frac{a\kpa^2 + b}{c \kpa^2 + d}\ d\kpa =
\frac{a}{c} + \frac{a}{c}\left(\frac{b}{a} - \frac{d}{c}\right)
\sqrt{\frac{c}{d}}\tan^{-1}\sqrt{\frac{c}{d}},
$$
$$
I_2 = \int_0^1 \frac{a\kpa^2 + b}{c \kpa^2 + d}\ \kpa^2\ d\kpa =
\frac{a}{3c} + \frac{a}{c}\left(\frac{b}{a} - \frac{d}{c}\right)
\left[ 1 - \sqrt{\frac{d}{c}}\tan^{-1}\sqrt{\frac{c}{d}} \right],
$$
and
$$
a = \left(\frac12 - i(\tp\op)\right) (1-u^2), \qquad
b = \frac12 (1+u^2) - i(\tp\op) u^2,
$$
$$
c = (\tp\op)^2 (1-u^2),\qquad
d = 1 + (\tp\op)^2 u^2.
$$
The real part of the expressions in Eq.~(\ref{eq:D_i}) should
be taken as respective components of the diffusion tensor.
For numerical stability purposes we perform the integration in
Eq.~(\ref{eq:D_i}) with the same precautions required for calculations of Yosida
functions. Namely, to avoid problems at $T \rightarrow 0$ and $\xi_k
\rightarrow 0$, we substitute the variable $\xi' = \tanh
(2\xi_k/7)$. If one is interested in the hydrodynamic limit $\ol\tau \ll
1$, then for $I_1$ and $I_2$ more numerically stable approximations are $I_1 =
a/3+b$ and $I_2 = a/5+b/3$.

Here we are interested in low temperatures, where $\ol \tau \gg 1$. At
these temperatures the values $|u| \ll 1$ are
important in the integrals in Eq.~(\ref{eq:D_i}). Thus in order to find a low-temperature approximation of $D$
we expand the integrands in Eq.~(\ref{eq:D_i}) in powers of $u$, using the definition
of Yosida functions
$$ Y_n(T) = \int_{-\infty}^{\infty} u^n \phi_k\ d\xi_k $$
to perform  the integration, and take the limit $\ol \tau \rightarrow
+\infty$. The resulting expressions, with expansion up to $u^4$, are
$$
D_{xx} =
\frac{v_F^2}{\ol \chi'}
\frac{\pi\lambda(2+\lambda)^3}{8(1+\lambda)^5}
\left(
Y_0
+ \frac12 (\lambda^2+2\lambda+4) Y_2
+ \frac38 [8+\lambda(2+\lambda)(4+2\lambda+\lambda^2)] Y_4
\right),
$$
$$
D_{zz} =
\frac{v_F^2}{\ol \chi'}
\frac{\pi(2+\lambda)^2}{4(1+\lambda)^5}
\left(
Y_0
- \frac12 (\lambda^2+2\lambda-2) Y_2
- \frac18 [-8+\lambda(2+\lambda)(8+2\lambda+\lambda^2)] Y_4
\right),
$$
where
$$
\chi' = \frac{2+Y_0}{3+F_0^a(2+Y_0)} \approx \frac{2}{3+2F_0^a},\qquad
\lambda = -F_0^a \chi'.
$$
Note that at least three terms in the expansion over Yosida functions are
needed to reach reasonable accuracy even at temperatures below $0.2 T_{\rm
  c}$. On the other hand, it is enough to keep the zeroth term of the expansion
over $(\ol\tau)^{-1}$ (i.e. the $\tau$-independent term) for satisfactory
calculation of $D$ at frequencies $\ol/2\pi \sim 1\,$MHz and temperatures
$T < 0.25 T_{\rm c}$. This has an important consequence for the thermometry.
Even if some magnetic relaxation is associated with the sample boundaries
and $\tau$ becomes smaller than in the infinite bulk liquid, the operation of
the magnon-condensate thermometer would not be affected provided that $\ol
\tau \gg 1$.

The exponential temperature suppression of $D$ comes from the behaviour of
Yosida functions at low temperatures,
$$
Y_n(T,\Delta) =
2\Gamma[(n+1)/2]
\left(\frac{T}{\Delta}\right)^{(n-1)/2}
\exp\left(-\frac{\Delta}{T}\right).
$$
For our pressure 4.1\,bar $F^0_a = -0.73$ and thus
$$
D_{xx} =
\frac{v_F^2}{\ol}
\left[
0.659
+2.234 \left(\frac{T}{\Delta}\right)
+19.90 \left(\frac{T}{\Delta}\right)^2
\right]
\sqrt{\frac{\Delta}{T}}
\exp\left(-\frac{\Delta}{T}\right),
$$
$$
D_{zz} =
\frac{v_F^2}{\ol}
\left[
 0.474
-0.185 \left(\frac{T}{\Delta}\right)
-11.71  \left(\frac{T}{\Delta}\right)^2
\right]
\sqrt{\frac{\Delta}{T}}
\exp\left(-\frac{\Delta}{T}\right).
$$
Note that in $^3$He-B $D_{xx} \ne D_{zz}$ in general. This
should be taken into account where necessary \cite{Fomin-anisD}. For example
Eq.~(\ref{eq:tau_SD_harm}) should be written as
$$
\tau_{\mathrm{SD}}^{-1} = \frac{65}{96}
  \frac{\omega_{\mathrm{L}}}{\xi^2_{\mathrm{D}}\Omega_B^2} (2D_{xx}\omega_r + D_{zz}\omega_z).
$$
In this work $\omega_z\ll\omega_r$ and therefore this correction is not important.

\bibliographystyle{plainnat}

\end{document}